# Women upskilling or reskilling to an ICT career: A systematic review of drivers and barriers

S. Williams[a], K. Blackmore[a], R. Berretta[a], and M. Mansfield[b].

[a] *School of Physical and Information Sciences, University of Newcastle, Newcastle, Australia;* [b] *Learning and Teaching, University of Newcastle, Newcastle, Australia.*

# Women upskilling or reskilling to an ICT career: A systematic review of drivers and barriers[1]

Demand for technology focused STEM professionals will increase globally over the coming decade, with many countries finding it difficult to meet growing demand. Compounding this are difficulties in attracting and retaining female technology-focused professionals. Research seeking to address this gender imbalance and workforce shortage focuses on increasing participation among school leavers. However, there is a paucity of research around the potential for females to upskill or reskill into an ICT career. As a starting point, this review asks the question: "What potential drivers and barriers have been identified that impact on female intentions or choices to reskill or upskill to a technology focused STEM career". Results indicate dissatisfaction in a first career, combined with positive computing experiences in the workplace can rouse interest in computing professions. Learning of job opportunities, especially from salient referents, is also a key driver. Results indicate women must overcome negative identity and academic beliefs, as well as self-doubt to make the switch. In summary, it is possible to increase and diversify the tech workforce by leveraging women's latent interest in computing. This review provides a roadmap for research to support educational institutions, employers, and women to benefit from upskilling or reskilling opportunities.

**Introduction**

Skill shortages in information and communications technology (ICT) areas are being felt globally. According to Statistica's data analysis (Sherif, 2023), the ICT industry faces skill shortages across several areas, particularly emerging technologies. This skills shortage is compounded by a lack of employee diversity, including gender diversity.

---

[1] The terms 'women' and 'females' are used interchangeably and without definition in the literature. For consistency, this review has followed the same format.

According to the World Economic Forum (WEF), less than 30% of the global ICT workforce is female (Pal et al., 2024). Addressing this gender disparity has been concentrated on school leaver initiatives. Central to these efforts is research seeking to understand drivers of career choice, aiming to funnel more females into tertiary ICT courses (Sharma et al., 2021). This review examines the drivers and barriers to increasing female participation in the ICT industry among non-school leavers, laying the groundwork for future research.

This review focusses on women upskilling or reskilling to ICT. Upskilling is defined as *individuals learning new skills to keep up with changes, boost performance, or position themselves for promotion within an existing career path* (Li, 2022). Reskilling is defined as *individuals acquiring new knowledge or skills to take on an entirely different role or career* (Li, 2022). The WEF (2020) emphasises the need for upskilling and reskilling to meet future job demands, as around 50% of an employee's current skill set can become obsolete within two years (Chifamba, 2020). A solution for skill obsolescence is continual skill development and career flexibility (Chifamba, 2020). The 2019 COVID pandemic underscored the need for workforce agility. Google Trends (2024) reveals a global uptick in searches for "upskilling" and "reskilling" from 1st January 2004 to 1st September 2024 (Figure 1), emphasising the need to identify factors influencing women's entry into ICT careers. Although this review focusses on ICT, it may application for STEM careers generally.

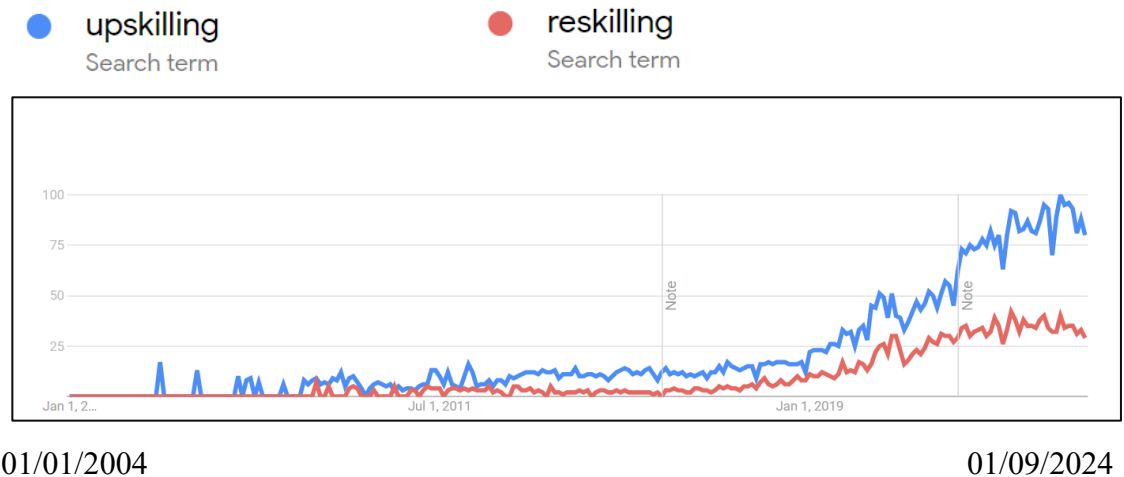

01/01/2004                                                                       01/09/2024

Figure 1 Interest in upskilling and reskilling (Google Trends, 2024)

**Background**

For this review, literature related to job or career change was considered to find a framework useful for synthesis and discussion. *The Integrated Career Change Model* proposed by Rhodes and Doering (1983) is a well-known model regularly used in job and career change research and discussions (Carless & Arnup, 2011; Uy, 2021). It provides a framework for identifying motivators for job or career change in the first portion of the model and covers the necessary processes to effect change in the second portion of the model. The full model is complex, with several elements less relevant to this study. This review relates to intentions or choices to change jobs or careers (first portion of the model); therefore, the actual process of job or career change (second portion of the model) can be disregarded. According to Rhodes & Doering (1983), the main motivator for job or career change is job or career dissatisfaction, with factors that cause this depicted in the model. Additionally, the model enables other unique factors (not related to job or career satisfaction) to initiate job or career change. To simplify the model, these factors are consolidated as: work environment factors, job performance and reward factors, personal factors, evaluation of current job outcomes versus alternative opportunities (job related factors), and other unique factors.

While the Rhodes and Doering model is useful, it lacks gender or ICT role specificity. There were no frameworks specific to ICT career change and advancement for women in the workforce located in the literature. Therefore, the framework will combine elements from the Rhodes and Doering generic career change model, with an adapted model on female perceptions of ICT majors by Gardner, Sheridan, and Tian (2014). While not a job or career change model, it includes factors that influence female first career choice in relation to ICT which could influence non-school leavers considering an ICT job or career change. The Gardner et al. model captures factors including job related beliefs, image related beliefs, academic related beliefs, experiential beliefs, self-evaluation, and salient referents. It is expected the literature will support these acknowledged drivers and barriers to ICT career choice for women.

Combining factors that motivate job or career change (Rhodes and Doering, 1983) with factors that influence female choice of ICT as a first career major (Gardner et al., 2014) provides a useful framework for this review. The merged and adapted framework is shown in Figure 2.

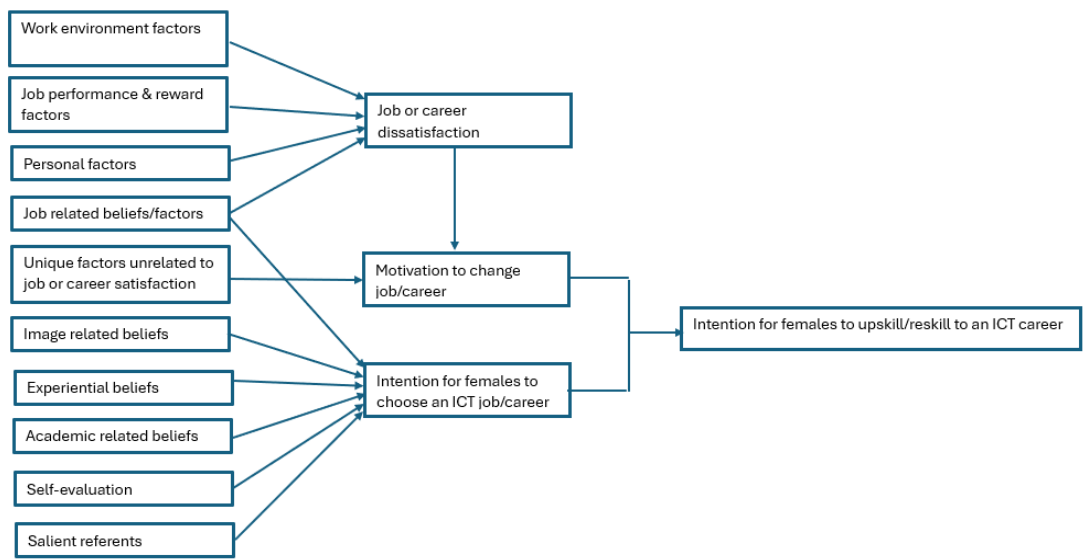

Figure 2. Proposed framework for 'Female Career Change to ICT' systematic review

The proposed Female Career Change to ICT (FCC2ICT) framework simplifies those dimensions relevant to female career change to ICT from both existing models and incorporates them into a single framework. The definitions for the factors in the framework are provided in the original literature source and are not explicitly defined here for conciseness. The purpose therefore of this systematic review is to begin the task of validating this framework by identifying existing literature that supports the inclusion of, and relationship between, factors.

**Methodology**

This study seeks to analyse and synthesise previous research on barriers and enablers influencing a decision to change career or upskill, where that research may be relevant to females exploring ICT professions. For this reason, the study has taken the form of a systematic review.

*Specification of Research Question*

To identify the most relevant literature, the following research questions were defined:

> RQ1: What **drivers** and/or **enablers** have been identified that might impact on female intentions or choices to reskill or upskill to a technology focused STEM career?
> RQ2: What **challenges** and/or **barriers** have been identified that might impact on female intentions or choices to reskill or upskill to a technology focused STEM career.

*Selection of Articles*

The research questions guide the development of an appropriate search strategy. The Preferred Reporting Items for Systematic Reviews and Meta-analyses (PRISMA) guidelines (Page et al., 2021) was implemented with adaptation for this study. The search strategy involved an electronic search of five key databases spanning business, computing, education, health, humanities, science, and social science.

Three search parameters were combined to focus the article search. The first parameter used search terms including (*upskill\* OR reskill\* OR retrain\* OR "career change"*) to exclude school leavers and focus on "in career" activities. The second used terms including (*tech\* OR ICT OR "information technology" OR comput\**) to identify articles with specific relevance to ICT. The final parameter used search terms (*gender OR female\* OR woman OR women*) to capture articles with a gender relevance.

Not all databases were able to use the same search and filtering options, with Boolean operators translated into the closest equivalent for each database. The inclusion period for this review was the 1st of January 2014 to the 14th of October 2024. As the research question is contemporary in nature, a ten-year time frame was chosen to exclude studies too dated for high relevance (Meline, 2006). All articles needed to be available in English. Sources were exported directly from each database to the referencing system Endnote (Clarivate, 2023).

An initial de-duplicating process was carried out in Endnote and further duplicates were identified on import to the Covidence platform (Veritas Health Innovation, 2023). A total of 1093 sources were subjected to a double-blind title and abstract screening process involving three (3) reviewers (Figure 3), with each source independently reviewed by two (2) reviewers. When reviewers disagreed on the inclusion status of a source, those reviewers met to discuss their line of reasoning regarding the agreed inclusion and exclusion criteria. The source's status was confirmed only when agreement was reached between the two reviewers. The inclusion and exclusion criteria used during the screening process is provided in Table 1.

**Table 1**: Inclusion and Exclusion Criteria

| Inclusion criteria: | Exclusion criteria: |
|---|---|
| Discusses career change, job change, upskilling or reskilling in the results or conclusions of the paper. | Females not included |
| | Career/job focus lacking |
| | ICT implication not found |
| Mentions and has findings relevant to ICT, computing, or digital careers. | Scholarly context lacking |
| | No full text available |
| | Not in English |

After this initial screening process, 968 sources were excluded. The remaining 125 sources were then subjected to independent full-text review by two (2) reviewers. Again, if there was disagreement on a source, the reviewers met to reach consensus. During full text review, 108 sources were excluded. The remaining 17 sources form the results of the PRISMA process and were exported for full analyses.

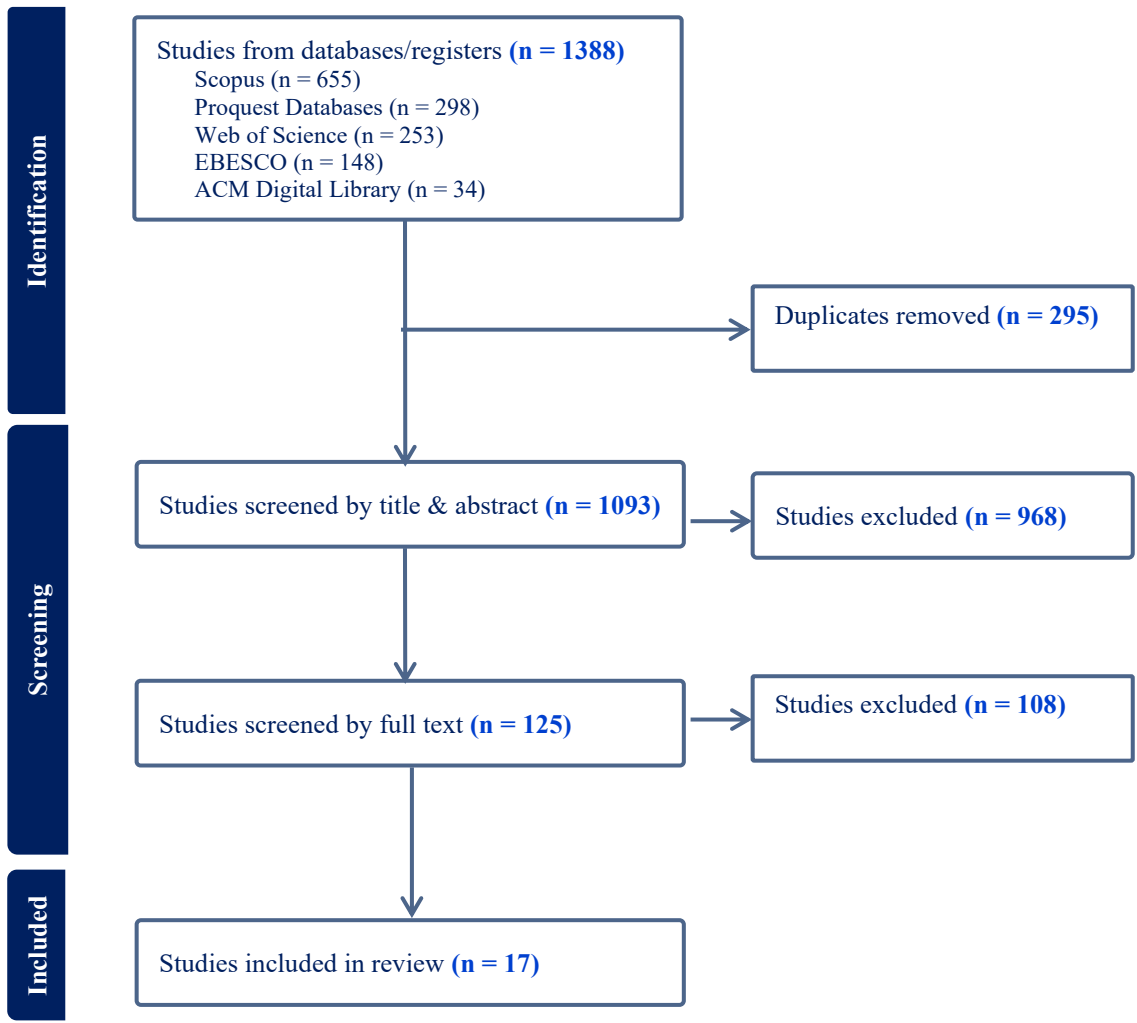

Figure 3: PRISMA screening diagram

*Analysis*

Most articles contained qualitative data based on interviews or focus groups. This data was predominately presented as quotes from participants, together with the conclusions reached by the researchers. When quantitative data was included, relevant findings were discussed in the paper. This review will therefore predominantly synthesise key findings from and in prose, utilising the ten factors from the combined framework to categorise results (Figure 2 & Table 2). As barriers can also act as enablers in the context of this review, a rating scale was developed to capture this duality (see Results).

**Results**

The results have been summarised in Table 2, across five columns. The first column lists the factors/beliefs from the FCC2ICT framework (Figure 2) and a brief description of the factor/belief. The second column counts the number of articles where the factor/belief is a driver and the number of articles where it is a barrier, using this to create the rating. The rating is calculated by taking the driver count and deducting the barrier count. Thus, a negative rating indicates the factor/belief was considered a barrier to women upskilling or reskilling to ICT in more of the articles. Conversely, a positive rating indicates it was considered a driver by more of the articles. This column also documents articles where the factor/belief was present. The third and final column lists all the *specific* drivers and barriers mentioned in the articles that sit within the framework factor/belief. A discussion of these results will then follow.

Table 2: Summary of barriers and enablers

| Factor/belief & Description | Specific drivers & barriers to upskill/reskill in ICT | Article count rating & Articles containing factor/belief |
|---|---|---|
| 1. Work environment factors<br><br>*Factors in the current work environment that effect job satisfaction, including organisation/ person & job/ person correspondence* | Drivers<br>• Boredom in current job/career<br>• Lack of opportunities in current career<br>• Insecure work in current job/career<br>• Long hours in current job/career<br>• Undesirable projects in current job/career<br><br>Barrier<br>• Current workplace not investing in upskilling or reskilling | Driver (4)<br>Barrier (1)<br>**Rating 3**<br><br>Brandi et al.(2023), Hyrynsalmi & Hyrynsalmi (2019b), Laine et al. (2023), Lyon & Green (2020), Ruthotto et al.(2021). |
| 2. Job performance and reward factors<br><br>*Correspondence between job performance and reward in current role* | Drivers<br>• Mismatch between pay level and work demands in current job/career | Driver (1)<br>Barrier (0)<br>**Rating 1**<br><br>Hyrynsalmi & Hyrynsalmi (2019b) |
| 3. Personal Factors<br><br>*The direct or indirect effect of personal* | Driver<br>• Growth in family commitments requires more flexible work options than current job/career offers<br><br>Barrier<br>• Young family makes it difficult to study for a career change | Driver (3)<br>Barrier (2)<br>**Rating 1**<br><br>Hyrynsalmi & Hyrynsalmi (2019a), |
| *characteristics on job satisfaction* | | Hyrynsalmi & Hyrynsalmi (2019b), Hyrynasalmi et al. (2024). |
| 4. Job related factors/beliefs ICT

*The effect beliefs about job/study opportunities have on job satisfaction and career choice* | Driver<br>- Expectation of high job availability<br>- High pay in job ads<br>- Social media posts showcasing career opportunities<br>- Expected meaningful & interesting work<br>- Expected remote or flexible work opportunities<br>- Expected career development opportunities<br>- Expected job security<br>- Expected equal pay for women<br>- Expected recognition for work<br>- Belief that experiences in other industries can be useful in an ICT career<br>- Real or perceived need to stay current in the ICT workforce<br>- Low entry bar for specific Masters degrees<br><br>Barrier<br>- Belief jobs in tech are boring<br>- Belief the ICT industry is too difficult | Driver (8)<br>Barrier (10)<br>**Rating  -2**<br><br>Aduragba et al. (2020), Billionniere & Rahman (2022a), Billionniere & Rahman (2022b), Brandi et al. (2023), Buhnova et al. (2019), Herman (2015), Hyrynsalmi (2019), Hyrynsalmi & Hyrynsalmi (2019a), Hyrynsalmi & Hyrynsalmi (2019b), Hyrynsalmi (2024), Hyrynsalmi et al. (2024), Laine et al. (2023), Lyon & Green (2020), Ruthotto et al. (2021), Thayer & Ko (2017). |

|  | - Belief gender equality unlikely to be obtained in ICT<br>- Belief mature workers unwelcome in the industry<br>- Real or perceived lack of opportunities for women returning to Science, Engineering and Technology (SET) careers<br>- Lack of ICT career change training information<br>- Second degree takes too long<br>- Lack of post graduate access for students with non-STEM undergraduate degrees |  |
|---|---|---|
| 5. Unique factors unrelated to job or career satisfaction<br><br>*Considers factors unrelated to job satisfaction or job/career choice* | Driver<br>- Unable to find work in current career<br>- Returning after a career break<br>- Goal of career advancement<br>- Seeking admission to PhD<br><br>Barrier<br>- Real or perceived ageism<br>- Real or perceived sexism | Driver (3)<br>Barrier (2)<br>**Rating 1**<br><br>Buhnova, Jurystova & Prikrylova (2019),<br>Billionniere & Rahman (2022b),<br>Herman (2015),<br>Ruthotto et al. (2021). |
| 6. Image related beliefs<br><br>*Image related beliefs are based on perceptions of careers and how these fit with an* | Driver<br>- None<br><br>Barrier<br>- Overcoming belief that women can't be programmers<br>- Overcoming belief that computing is a male's domain | Driver (0)<br>Barrier (8)<br>**Rating -8**<br><br>Billionniere & Rahman (2022a),<br>Billionniere & Rahman (2022b),<br>Buhnova et al. (2019), |
| *individual's personal image* | • Overcoming lack of encouragement to pursue ICT as a female<br>• Concerns of poor treatment in a male heavy industry<br>• Imposter syndrome<br>• Lack of support building resumes and preparing for interviews | Hyrynsalmi (2019), Hyrynsalmi & Hyrynsalmi (2019b), Hyrynsalmi (2024), Lyon & Green (2020), Thayer & Ko (2017). |
| *7. Academic related beliefs*<br><br>*Refers to how difficult an individual perceives a discipline area to be, and how the individual's learning style conforms to anticipated teaching styles in that discipline* | Driver<br>• Low bar for entry to specific Masters degree<br>• Flexible learning options<br>• Access to preparatory courses<br>• Courses that were all female or encouraged and catered for diversity<br>Barrier<br>• Negative view of STEM education<br>• Assumption high level math skills are required or fail to meet math pre-requisites<br>• Assumption programming knowledge pre-requisite for reskilling in computing<br>• Difficult finding right combination of online and face-to-face training options<br>• Concerns around keeping up with the pace of technology, especially after a career break<br>• Predominance of male students in learning environments with curriculum and pedagogy more suited to males | Driver (4)<br>Barrier (7)<br>**Rating -3**<br><br>Billionniere & Rahman (2022b),<br>Buhnova et al. (2019),<br>Hyrynsalmi & Hyrynsalmi (2019a),<br>Hyrynsalmi & Sutinen (2019),<br>Hyrynsalmi et al. (2024),<br>Laine et al. (2023),<br>Lyon & Green (2020),<br>Thayer & Ko (2017). |
| 8. Experiential beliefs<br><br>*This refers to the level of mental reward anticipated by an activity* | Driver<br>• Positive experience with technology in the workplace<br>• Positive experience in all female tech groups<br><br>Barrier<br>• Negative ICT school experiences<br>• Lack of exposure to ICT in high school<br>• Lack of information and support around ICT as career choice<br>• Negative experiences working in the computing industry<br>• Negative experience in male dominated tech social groups | Driver (5)<br>Barrier (5)<br>**Rating 0**<br><br>Billionniere & Rahman (2022a),<br>Hyrynsalmi (2024),<br>Hyrynsalmi et al. (2024),<br>Laine et al. (2023),<br>Lyon & Green (2020),<br>Lyon & Green (2021),<br>Ruthotto et al. (2021). |
| 9. Self-evaluation<br><br>*Self-evaluation refers to an individual's perception of their skills or ability to learn new skills and be successful in a specific area* | Driver<br>• None<br><br>Barrier<br>• Concerns around keeping up with technology<br>• General self-doubt<br>• Lack of confidence in what to learn or ability to make the career change<br>• Low self-esteem in job hunting | Driver (0)<br>Barrier (2)<br>**Rating -2**<br><br>Hyrynsalmi (2019),<br>Hyrynsalmi & Hyrynsalmi (2019a). |
| 10. Salient referents | Driver<br>• Partner working in the computing industry<br>• Family support to enter training | Driver (3)<br>Barrier (3)<br><br>**Rating 0** |

| *Salient referents are those important enough to influence an individual's career choice* | • Friends working securely in the industry<br><br>Barrier<br><br>• Family seeing computing as the prevue of males<br><br>• Friends not seeing computing as a good choice | Hyrynsalmi & Hyrynsalmi (2019b), Hyrynsalmi et al. (2024), Laine et al. (2023), Lyon & Green (2020), Thayer & Ko (2017). |
|---|---|---|

**Discussion**

This section will further explore the qualitative data found in review literature, providing a richer picture of specific barriers and enablers within the ten factors of the FCC2ICT framework.

(1) Work environment factors

Work environment factors include aspects and provisions of current employment such as working hours, type of projects, and pay (Lyon & Green, 2020). Work environment factors can act as a driver or barrier for upskilling and reskilling in ICT. It is worth noting this factor has the highest positive rating score (3) in the results, indicating work environment factors are more likely to act as drivers. When negative work environment factors accumulate, they can produce a strong incentive for women to consider job or career change. In most cases, job dissatisfaction combines with other factors to initiate upskilling or reskilling to ICT. Boredom with early career choice can combine with positive experiential ICT beliefs (Lyon & Green, 2020). A lack of career opportunities along a first career path, can combine with confidence that there are multiple, stable job opportunities in ICT (Hyrynsalmi & Hyrynsalmi, 2019b). In summary, perceived negative work environment factors are important drivers of job/career change, especially when combined with other factors in the FCC2ICT model.

(2) Job performance and reward factors

Job performance and reward factors were identified by only one resultant study, with Hyrynsalmi & Hyrynsalmi (2019b) finding a mismatch between work demands and pay level can lead to career dissatisfaction and career change. For example, in the study by Hyrynsalmi & Hyrynsalmi (2019b), a female participant indicated that pay in the healthcare sector did not match high-level demands, which led them to consider a well-paid ICT career. Overall, this factor had a low positive rating score (1) indicating job dissatisfaction combined with positive job-related beliefs acts as a driver for ICT career change.

(3) Personal factors

Personal factors include the direct and indirect effect of personal characteristics on job satisfaction. Family commitments can be both a driver toward an ICT career and a challenge for any career change. On the one hand, family was seen as a motivator to change careers to the software industry, as it was perceived to have a more flexible work environment (Hyrynasalmi, et al. 2024). Alternatively, family commitments can make it difficult for women to undertake career change studies (Hyrynsalmi & Hyrynsalmi, 2019b). Personal factors have a low positive rating score (1) and indicate these can be drivers, but support is required to facilitate ICT career change.

(4) Job related factors/beliefs

The real, or perceived, availability of jobs in ICT is an important driver of career change (Aduragba, 2020). Lyon & Green (2020) report that advertisements suggesting lucrative jobs are available in the software industry motivated several women to change careers. Generally, women expected to find meaningful and interesting work, remote work opportunities, and good flexibility in the ICT sector (Laine et al., 2023). Women who wished to upskill or reskill via a Masters in IT also expected good career development opportunities and work flexibility after graduating (Laine et al., 2023). The literature indicated that generally women were successful at upskilling or career change to ICT. However, for some boot camp graduates it was difficult to secure employment upon graduation (Lyon & Green, 2021).

Beliefs about what it would be like to work in ICT were motivators for some women. High wages, equal pay for women, and opportunity to take on societal challenges were motivators (Hyrynsalmi & Hyrynsalmi, 2019a). Others reported interest in new technologies and varying job opportunities as drivers for their career change (Hyrynsalmi & Hyrynsalmi, 2019a). Hyrynasalmi (2024) found enthusiasm for technology and its meaningful application is important for women to choose a tech career.

There were also a range of negative beliefs about work in the tech industry. Many women believe tech jobs are too boring, or too difficult, due to lack of exposure (Buhnova et al., 2019). Some women indicated their desire for gender equality in the workplace was unlikely to be realized within computing professions (Laine et al., 2023). Still other women were more concerned about age discrimination than gender bias, questioning if the industry preferred recruits in their 20s (Hyrynsalmi, 2019).

Hyrynsalmi & Hyrynsalmi (2019b) reported that women with existing non-technical degrees believed the software industry might allow them to capitalise on their knowledge and experience because software is used across industries. For women already in ICT, enrolment in further education is linked to intrinsic motivation, staying relevant in the workforce, and maintaining a current role (Ruthotto et al., 2021). For those returning after a career break, upskilling gave them confidence to apply for jobs (Herman, 2015).

According to the Rhodes and Doering (1983) model, job related beliefs also extend to training opportunities and the financial capacity to capitalise on those opportunities.  Hyrynsalmi and Hyrynsalmi (2019a) found a lack of course comparison information and advanced short courses were challenges to career change to the software industry. Microsoft and Codeacademy were among the popular short course choices for those wanting to start in this industry (Hyrynasalmi, 2024). Bootcamps were also popular, but a lack of finance and dedicated study time were barriers for some (Lyon & Green, 2020).

It has been proposed that having additional short, skill-based programs, and more online options would assist with upskilling and reskilling women in tech (Billionniere & Rahman, 2022b). However, some women emphasise the importance of face-to-face teaching and community on the journey of ICT career change . Masters degrees in ICT were considered by women with non-tech degrees, however, they were generally ineligible to enroll in them (Hyrynsalmi & Hyrynsalmi, 2019a). ICT Masters with low bar entry requirements were an exception, attracting female enrolments (Hyrynasalmi, et al. 2024).

In summary, although positive job-related beliefs have multiple drivers, negative job-related beliefs that create barriers are found across more review articles. These negatives beliefs can reduce the impact of positive job-related beliefs. Therefore, job-related beliefs fall into the barrier category, with a low negative rating score (-2).

(5) Unique factors unrelated to job or career satisfaction

Various reasons were identified for women reskilling or upskilling in ICT that were not directly related to job or career satisfaction. Some women were attending tech training courses because they could not find jobs in their chosen field after they graduated (Buhnova et al., 2019). Others were upskilling after having taken time out from their ICT career to have a family (Billionniere & Rahman, 2022b). Other motivators for women to upskill included career advancement and to attain a PhD (Ruthotto et al, 2021).

There were also some barriers to upskilling in ICT unrelated to job or career satisfaction. Women seeking to return to STEM careers after a break cited sexism and ageism as factors that dampened their desire to reskill and upskill to re-enter the profession (Billionniere & Rahman, 2022b). Research suggests there are a set of gendered assumptions in STEM workplaces affecting culture and recruitment practices that make returning to difficult or less appealing to women (Herman, 2015).

Although there are negative unique factors unrelated to job or career satisfaction, overall unique factors tend to be drivers, resulting in an overall low positive score (1).

(6) Image related beliefs

Image related beliefs is one of two factors in this review with no identified drivers. Image related beliefs were frequently noted as barriers, bringing the rating to a very high negative (-8). Image related beliefs were cited as being barriers to ICT as a first career choice, and something that had to be challenged to consider a tech-focused career change (Thayer & Ko, 2017). Some participants who attended computer classes during university were discouraged by stereotypes, the masculine nature of the industry, lack of incentive to ask questions, and feelings of imposter syndrome (Lyon & Green, 2020). Image related beliefs can continue to be obstacles for mature age women considering upskilling or reskilling in computing. Some women reported struggling to see themselves as qualified, or convince others they were qualified, when returning from a career break or entering from a non-STEM background (Billionniere & Rahman, 2022; Hyrynsalmi & Hyrynsalmi, 2019b). Some females felt their male peers were much more willing to 'fake it till they make it' when applying for jobs for which they did not have all the requirements (Billionniere & Rahman, 2022b). Women also expressed concerns about working in male dominated environments (Buhnova et al., 2019; Thayer & Ko, 2017). Some had experienced condescension or a lack of trust in their abilities from male colleagues (Hyrynsalmi, 2019). However, some women believe attitudes to women are improving, and a small number prefer to work in a male-heavy industry (Hyrynsalmi, 2019).

(7) Academic related beliefs

Academic related beliefs can be barriers to women considering computing careers. According to Social Cognitive Career theory (Lent et al., 1994), a person's inputs into a learning experience affect career plans; therefore, women have more difficulty entering computer science majors due to limited background and personal inputs that support their entry (Lyon & Green, 2021). Indeed, negative academic beliefs continue as barriers for career changers. Women tend to have negative perceptions of STEM education, particularly math and science components (Billionniere & Rahman, 2022b). Women often assume software engineers need deep level math, and that coding skills are essential before enrolling in coding courses; such beliefs deter career change (Hyrynasalmi, et al., 2024). Often these barriers are re-enforced by education providers through prerequisites, making ICT study prohibitive for some women (Lyon & Green, 2020; Hyrynsalmi & Hyrynsalmi, 2019b).

From the studies, women also expressed negative academic beliefs about upskilling and maintaining employment. They were concerned about the fast-moving nature of technology and how to keep up without support, particularly during a career break (Billionniere & Rahman, 2022b).

Another issue influencing negative academic beliefs was the predominance of males in ICT learning environments. Many bootcamps, for example, had very few women (Thayer & Ko, 2017). Research suggests masculine culture can contribute to a lower sense of belonging in female STEM students (Fisher et al., 2020).

However, there were bootcamps that strived for greater diversity, and some that only accepted female students (Thayer & Ko, 2017). Brandi et al (2023) suggests it is important courses are inclusive and provide a range of perspectives and materials in their learning design to increase student diversity. Indeed, review-based research recommends gendered learning preferences be considered to support female STEM students (Fisher et al., 2020). Workplaces can also provide a supportive environment to increase diversity (Brandi et al., 2023).

When it comes to educational delivery modes, a mix of online and face-to-face produced more positive academic beliefs in women. However, the order and mix of these mattered. Some women felt online courses were good for getting started, but more guidance was preferred as learning progressed (Hyrynsalmi & Hyrynsalmi, 2019b). Conversely, Buhnova et al. (2019) found many women need more guidance when getting started in tech. Lyon and Green (2020) found many women that enrolled in bootcamps had tried to teach themselves to code but found this very difficult and slow. The women were looking for a structured learning environment with an expediated graduation timeline (Lyon & Green, 2020).

Post graduate degrees in computing with easier or supported entry were an academic confidence builder for women. Laine et al. (2023) found the top reasons for enrolling in a Masters in IT were centered around flexibility to learn online and obtain entry without an entrance exam. These motivators were cited ahead of actual interest in ICT (Laine et al., 2023).

In summary, academic related beliefs can be drivers or barriers. However, they are more often cited as barriers in the review literature, with a low negative rating (-3).

(8) Experiential beliefs

Lack of exposure to computing in schools is a barrier to women entering computing majors straight from school (Lyon & Green, 2020). Some women attempting career change to the software industry had no exposure to programming until after college, others had some school exposure, but too late to choose computer science as a major (Lyon & Green, 2020). Lyon and Green (2021) found in their study that women who majored in computer science had earlier introductions to programming than women who had enrolled in boot camps as career changers. Ruthotto et al. (2021) found that most female reskillers and upskillers who enrolled in an online Master of Computer Science, did not have a computing undergrad degree. However, the majority did have some experience working in a computing job of some kind (Ruthotto et al., 2021).

Indeed, a positive experience with computing in the workplace can be a catalyst for career change later in life (Lyon &Green, 2020; Hyrynasalmi, et al., 2024). Being exposed to the work of ICT professionals in an organizational setting, or working on projects that include ICT elements, can raise interest in ICT as a second career choice (Hyrynasalmi, et al. 2024). For example, women can discover an enjoyment of programming while automating a work-related task (Lyon & Green, 2020). If such interest is roused, career and job guidance are very important enablers for women considering a tech focused career change (Hyrynasalmi, 2024)

Community is another place where women's ICT experiential beliefs are formed. Thayer and Ko (2017) found negative experiences in the male dominated gaming culture can be a deterrent to some women considering ICT professions. However, supportive mixed gender tech meetup groups, or women only tech communities can help combat this (Thayer & Ko, 2017; Hyrynasalmi, 2024).

It should be noted that not all women who have worked in the computing industry have positive experiential beliefs about the profession. Billionniere (2022) found the main reasons women are reluctant to re-enter computing after a career break are stereotypes in the industry, lack of returnships, and lack of supportive networks.

Positive and negative experiential beliefs are mentioned evenly in the review literature, providing a neutral rating (0).

(9) Self-evaluation

Studies indicate females experience higher levels of self-doubt than males when considering computing as a first career (Fisher et al., 2020). Self-doubt continues as an obstacle to women upskilling or reskilling to tech roles. Women express concern over their ability to learn the necessary skills needed for career change, and their ability to keep up with evolutions in the industry once there (Hyrynsalmi, 2019). Even after successfully completing courses to prepare themselves for the software industry, low self-esteem caused female job hunters to hesitate when applying for jobs because they did not possess all the attributes requested in job advertisements (Hyrynsalmi & Hyrynsalmi, 2019a). It seems many women making a career change to the software industry can doubt their own ability and potential for success (Hyrynsalmi, 2019).

Self-evaluation is one of only two factors from the existing frameworks that inform this review to have no drivers in the review literature. It is a factor that is not widely explored in the studies and warrants further research with regard to female career changers. Self-evaluation attracts a low negative rating (-2).

(10)   Salient referents

Family influence can be a barrier or a driver for ICT career change. Family indicating computing is for 'guys', for example, can be a deterrent (Lyon & Green, 2020). Women report that family members can struggle to understand why a tech career is even being considered (Thayer & Ko, 2017). Family may also encourage pursuit of areas such as business studies, over computing (Hyrynasalmi, et al. 2024). On the other hand, female ICT career changers can be motivated by a partner already successfully working in the industry (Hyrynsalmi & Hyrynsalmi, 2019b). In addition, family plays a key role in providing the support necessary for ICT career transition (Laine et al., 2023).

Likewise, the opinion of friends can be a barrier (Thayer & Ko, 2017) or driver to ICT career change (Lyon & Green, 2020). Several career change women were influenced by their friends who were already happily working in the software industry (Lyon & Green, 2020). Women are often motivated by friends who have personal or second-hand stories of successful career change (Lyon & Green, 2021). Hyrynsalmi and Hyrynsalmi (2019b) conclude that women's close circle can be leveraged to entice career change to the tech industry.

In conclusion, salient referents act both as enablers and barriers to women re-upskilling in ICT with an overall neutral (0) rating.

**Conclusion**

This review combined a career change framework, with a model predicting female choice as an ICT major in a first career. The literature supports the resulting FCC2ICT framework, confirming a range of drivers and barriers for women fit within the framework. The driver with the highest rating was negative work-related experiences in a first career choice. These negative work-related experiences, when combined with an ICT career choice driver, influence a decision to upskill or reskill in ICT. Prominent ICT career choice drivers were positive work-related computing experiences, and positive beliefs about computing jobs. Women tend to encounter these positive ICT experiences later in life. Prominent barriers to women choosing ICT as a second career were image-related beliefs, academic-related beliefs, and low self-evaluation. This is consistent with women not choosing ICT as a first career. Educational institutions can capitalise on drivers and address women's low self-confidence with a range of study options. These could include on-campus and face-to-face mixed mode offerings, low bar or supported entry programs, consideration of STEM gendered learning styles, and encouragement for women from non-STEM backgrounds to apply. Businesses can capitalise on drivers and lower barriers by supporting female employees looking to upskill or reskill in computing fields, and encouraging a culture that values mature, female participation in computing divisions. Further research will pursue best practice for recruiting and supporting women who may consider leaping into a technology future.

**Limitations and Future Work**

Although this review is based on seventeen unique articles, there are only ten unique first authors, which could result in some duplication of emphasis. The majority of articles focus on reskilling in software engineering, rather than the broader computing industry. Also, over seventy-five percent of the studies were conducted in the US or Finland. Studies focused on women who had started the upskilling or reskilling process, thus excluding women who had chosen not to upskill or reskill in ICT, leaving a critical gap in the literature that requires attention. Lastly, the literature contained several interventions for women who had chosen to upskill or reskill. While these interventions are beyond the scope of this review, this again provides a fertile direction for future research. Significant effort is directed to programs and interventions aimed at increasing female participation in ICT study and careers. This review highlights the barriers and enablers facing women seeking to upskill or reskill to ICT, and provides important insights to guide the development of educational or employer based interventions targeting women in non-ICT career areas.

**Declaration of interest**

No potential conflict of interest was reported by the authors

**References**

Aduragba, O. T., Yu, J., Cristea, A. I., Hardey, M., & Black, S. (2020). Digital inclusion in Northern England: Training women from underrepresented communities in tech: A data analytics case study. *15th International Conference on Computer Science & Education*, 162–168. https://doi.org/10.1109/ICCSE49874.2020.9201693

Billionniere, E., & Rahman, F. (2022a). Bridging the gender and skills gaps with emerging technologies. *Proceedings of the ASEE Annual Conference & Exposition*, 1-6.

https://search.ebscohost.com/login.aspx?direct=true&db=asn&AN=172835785&site=ehost-live

Billionniere, E., & Rahman, F. (2022b). Women of color in emerging technology: Breaking down the barriers. *Proceedings from ASEE Annual Conference and Exposition*, Minneapolis, US. https://peer.asee.org/40515,

Brandi, U., Hodge, S., Hoggan-Kloubert, T., Knight, E., & Milana, M. (2023). The European year of skills 2023: Skills for now and in the future? [Editorial]. *International Journal of Lifelong Education*, *42*(3), 225-230. https://doi.org/10.1080/02601370.2023.2212424

Buhnova, B., Jurystova, L., & Prikrylova, D. (2019). Assisting women in career change towards software engineering: Experience from Czechitas NGO. *Proceedings from the 13th European Conference on Software Architecture*, Paris, France. https://doi.org/10.1145/3344948.3344967

Carless, S., & Arnup, J. (2011). A longitudinal study of the determinants and outcomes of career change. *Journal of Vocational Behavior*, *78*(1), 80-91. https://doi.org/10.1016/j.jvb.2010.09.002

Chifamba, C. (2020). Career flexibility: A panacea to skills obsolescence. *Asian journal of education and social studies*, *7*(4), 12-16.

Clarivate. (2023). *Endnote* (Version 21) [Computer software]. Clarivate. https://endnote.com/

Fisher, C. R., Thompson, C. D., & Brookes, R. H. (2020). Gender differences in the Australian undergraduate STEM student experience: A systematic review. *Higher Education Research & Development*, *39*(6), 1155-1168. https://doi.org/10.1080/07294360.2020.1721441

Gardner, L.A., Sheridan, D., & Tian, X.E. (2014). Perceptions of ICT: An exploration of gender differences. *Proceedings from EdMedia 2014 World Conference on Educational Media and Technology* (pp. 120-129). Tampere, Finland. https://www.researchgate.net/publication/277516974_Perceptions_of_ICT_An_Exploration_of_Gender_Differences

Google Trends (2024). "Upskilling" and "reskilling", Retrieved September 1, 2024, from https://trends.google.com/trends/explore?date=all&q=upskilling,reskilling

Herman, C. (2015). Returning to STEM: Gendered factors affecting employability for mature women students. *Journal of Education & Work*, *28*(6), 571-591. https://doi.org/10.1080/13639080.2014.887198

Hyrynsalmi, S. (2019). *The underrepresentation of women in the software industry: Thoughts from career-changing women.* IEEE/ACM 2nd International Workshop on Gender Equality in Software Engineering. Montreal, Canada. https://doi.org/10.1109/GE.2019.00008

Hyrynsalmi, S., Sutinen, E. (2019). *The role of women software communities in attracting more women to the software industry*. International ICE Conference on Engineering Technology and Innovation. Nice, France. https://ieeexplore.ieee.org/document/8792673

Hyrynsalmi, S., & Hyrynsalmi, S. (2019a). *Software engineering studies attractiveness for the highly educated women planning to change career in Finland*. IEEE/ACM 41st International Conference on Software Engineering. Montreal, Canada. https://ieeexplore.ieee.org/document/8802816

Hyrynsalmi, S., & Hyrynsalmi, S. (2019b). *What motivates adult age women to make a career change to the software industry?* International ICE Conference on

Engineering Technology and Innovation. Nice, France. https://doi.org/10.1109/ICE.2019.8792630

Hyrynsalmi, S. M. (2024). *Rebooting the system and building new futures: Supporting women's comeback in IT*. 5th ACM/IEEE Workshop on Gender Equality, Diversity, and Inclusion in Software Engineering. Lisbon, Portugal. https://doi.org/10.1145/3643785.3648485

Hyrynsalmi, S. M., Peltonen, E., Vainionpää, F., & Hyrynsalmi, S. (2024). *The Second round: Diverse paths towards software engineering*. 5th ACM/IEEE Workshop on Gender Equality, Diversity, and Inclusion in Software Engineering. Lisbon, Portugal. https://doi.org/10.1145/3643785.3648494

Laine, S., Målquist, U., Myllymäki, M., & Hakala, I. (2023). *Women in IT project: Survey results*. 32nd Annual Conference of the European Association for Education in Electrical and Information Engineering, Eindhoven, Netherlands. https://doi.org/10.23919/EAEEIE55804.2023.10181717

Lent, R. W., Brown, S. D., & Hackett, G. (1994). Toward a unifying social cognitive theory of career and academic interest, choice, and performance. *Journal of Vocational Behavior, 45*(1), 79-122. https://doi.org/10.1006/jvbe.1994.1027

Li, L. (2022) Reskilling and upskilling the future-ready workforce for Industry 4.0 and beyond. *Information Systems Frontiers*. https://doi.org/10.1007/s10796-022-10308-y

Lyon, L. A., & Green, E. (2020). Women in coding boot camps: An alternative pathway to computing jobs. *Computer Science Education*, *30*(1), 102-123. https://doi.org/https://doi.org/10.1080/08993408.2019.1682379

Lyon, L. A., & Green, E. (2021). Coding boot camps: Enabling women to enter computing professions. *Acm Transactions on Computing Education*, *21*(2). https://doi.org/https://doi.org/10.1145/3440891

Meline, T. (2006). Selecting studies for systemic review: Inclusion and exclusion criteria. *Contemporary Issues in Communication Science and Disorders*. 33. 21-27. 10.1044/cicsd_33_S_21.

Page, M, McKenzie, J., Bossuyt, P., *et al.* The PRISMA 2020 statement: An updated guideline for reporting systematic reviews. *Systematic Reviews 10*(1), 89. https://doi.org/10.1186/s13643-021-01626-4

Pal, K., Piaget, K., & Zahidi, S. (2024). Global gender gap report 2024. *World Economic Forum*. https://www.weforum.org/publications/global-gender-gap-report-2024

Rhodes, S. R., & Doering, M. (1983). An integrated model of career change. *The Academy of Management Review*, *8*(4), 631–639. https://doi.org/10.2307/258264

Ruthotto, I., Kreth, Q., & Melkers, J. (2021). Entering or advancing in the IT labor market: The role of an online graduate degree in computer science. *Internet and Higher Education*, *51*. https://doi.org/10.1016/j.iheduc.2021.100820

Sharma, K., Torrado, J. C., Gómez, J., & Jaccheri, L. (2021). Improving girls' perception of computer science as a viable career option through game playing and design: Lessons from a systematic literature review. *Entertainment Computing*, *36*. https://doi.org/10.1016/j.entcom.2020.100387

Sherif, A. (2023). *IT skill shortages faced by IT leaders worldwide 2021*.Statistica. https://www.statista.com/statistics/662423/worldwide-cio-survey-function-skill-shortages/

Thayer, K., & Ko, A. (2017, Aug 18-20). Barriers faced by coding bootcamp students. In *Proceedings of the 2017 ACM Conference on International Computing Education Research*, Tacoma, WA.

Uy, Jeffrey. (2021). Determinants of career change: A literature review. JPAIR Multidisciplinary Research. 42. 1-19. 10.7719/jpair.v42i1.804.

Veritas Health Innovation. (2023). *Covidence* [Computer software]. Veritas Health Innovation. www.covidence.org

World Economic Forum. (2020). *The future of jobs report 2020*. https://www3.weforum.org/docs/WEF_Future_of_Jobs_2020.pdf